\RequirePackage{fix-cm}
\documentclass[twocolumn]{svjour3}          
\smartqed  
\usepackage{amsmath}
\usepackage{graphicx}
\usepackage[table,pdftex,dvipsnames]{xcolor}

\usepackage{url} 
\usepackage[hidelinks]{hyperref}

\usepackage{lipsum}                     
\usepackage{xargs}                      
\usepackage[colorinlistoftodos,prependcaption,textsize=tiny]{todonotes}
\newcommandx{\unsure}[2][1=]{\todo[linecolor=red,backgroundcolor=red!25,bordercolor=red,#1]{#2}}
\newcommandx{\change}[2][1=]{\todo[linecolor=blue,backgroundcolor=blue!25,bordercolor=blue,#1]{#2}}
\newcommandx{\info}[2][1=]{\todo[linecolor=OliveGreen,backgroundcolor=OliveGreen!25,bordercolor=OliveGreen,#1]{#2}}
\newcommandx{\improvement}[2][1=]{\todo[linecolor=Plum,backgroundcolor=Plum!25,bordercolor=Plum,#1]{#2}}
\newcommandx{\thiswillnotshow}[2][1=]{\todo[disable,#1]{#2}}
\begin{document}

\title{Unveiling the Digital Fingerprints
}
\subtitle{Analysis of Internet attacks based on website fingerprints}



\author{Blerim Rexha         
        \and
        Arbena Musa 
        \and
        Kamer Vishi
        \and
        Edlira Martiri
}


\institute{Blerim Rexha, Arbena Musa \at
              University of Prishtina \\
              Faculty of Electrical and Computer Engineering \\
              Kodra e Deillit pn 1000 Prishtina, Kosovo \\
              \email{arbena.musa@uni-pr.edu}           
           \and
            Kamer Vishi \at
            University of Oslo \\
            Department of Informatics, Oslo, Norway  
            \and
            Edlira Martiri \at
            University of Tirana \\
            Faculty of Economics, Tirana, Albania\\
}

\date{Published online: 2024}

\maketitle

\begin{abstract}
Parallel to our physical activities our virtual presence also leaves behind our unique digital fingerprints, while navigating on the Internet. These digital fingerprints have the potential to unveil users' activities encompassing browsing history, utilized applications, and even devices employed during these engagements. Many Internet users tend to use web browsers that provide the highest privacy protection and anonymization such as Tor. The success of such privacy protection depends on Tor feature to anonymize end user IP address and other metadata that constructs the website fingerprint. In this paper, we show that using the newest machine learning algorithms an attacker can deanonymize Tor traffic by applying such techniques. In our experimental framework, we establish a baseline and comparative reference point using a publicly available dataset from Universidad Del Cauca, Colombia. We capture network packets across 11 days, while users navigate specific web pages, recording data in .pcapng format through the Wireshark network capture tool. Excluding extraneous packets, we employ various machine learning algorithms in our analysis. The results show that the Gradient Boosting Machine algorithm delivers the best outcomes in binary classification, achieving an accuracy of 0.8363. In the realm of multi-class classification, the Random Forest algorithm attains an accuracy of 0.6297.


\end{abstract}

\section{Introduction}
\label{intro}

In today's interconnected world, the Internet has significantly influenced the formation of virtual identities during users' daily activities. Our digital identities are shaped not only by the intentional information we provide, but also by the vast amount of digital traces generated through our interconnected actions. These digital traces serve as identifiers and can be used to track our online activities, leading to potential privacy violations and exploitation. The continuous technological advancements and increasing reliance on the Internet have further exacerbated these privacy concerns, creating a unique level of invasion. The Google claims that about 95\% of the web traffic is encrypted towards their services, and that today 97\% of the web sites use HTTPS \cite{HTTPS} protocol, this still leaves behind digital traces, now known as digital fingerprints, exposing some user information.



To safeguard their privacy, many Internet users resort to anonymous networks such as Tor to conceal their source and destination IP addresses when accessing online services \cite{kovalchuk2021dark}. However, even within these networks, a threat known as Website Fingerprinting Attack (WF) persists \cite{liu2023survey}. Through this attack, an eavesdropping party can observe network traffic data, including the volume, size, and timing of data transmissions. Anonymous networks, although designed to preserve privacy, are also exploited by criminals to conceal illegal activities \cite{Chaabane2010}, hamper government forensic efforts, and increase monitoring capabilities\cite{Cadena2020}.

This delicate balance between protecting privacy and monitoring illicit and anonymous activities raises important questions regarding user surveillance practices.

Furthermore, as the role of machine learning continues to expand, it has emerged as a promising trend in the field of cybersecurity. Machine learning techniques facilitate the collection and analysis of sophisticated data to develop effective algorithms, detect patterns, and predict and respond to real-time cyber attacks. A comprehensive review of machine learning supported threat and attack detection is presented in \cite{MOUSTAFA2019,GIBERT2020}. Furthermore, machine learning algorithms could be applied to biometric authentication, as presented by \cite{Arbena}, combined with mobile devices as in \cite{Blerim} or to enhance the feature list of standard security tools, such as Burp Suite, as explained by \cite{Rrezearta}.

By enabling cybersecurity systems to understand threat patterns and learn cybercriminal behaviors, machine learning empowers proactive defense, reduces reliance on experts for routine tasks, and helps prevent future attacks \cite{Cadena2020}.


\section{Related Work}
\label{related_work}

\subsection{Network Traffic Monitoring}
\label{network_traffic_monitoring}
Monitoring network traffic is a crucial task in ensuring the security and smooth operation of computer networks. As networks grow increasingly complex and large, monitoring and managing them pose significant challenges. Consequently, researchers have dedicated efforts to develop diverse tools and techniques for monitoring and analyzing network traffic, identifying potential threats, and optimizing network performance \cite{Wireshark}. Several studies have reviewed prominent approaches in Internet network monitoring, including network traffic analysis\cite{Siswanto2019}. These reviews delve into the advantages, limitations, current trends, and research and development prospects in the field.

Early on, network traffic monitoring emerged as a strategic component in understanding and characterizing user activities, as highlighted in \cite{Finamore2011}. The study in \cite{Dorado2012} also observed an increased research focus on network monitoring, recognizing its pivotal role in comprehending the Internet and its users. Some studies have delved into comprehensive characterizations of specific applications or networks, considering perspectives from Internet service providers (ISPs) or end-users.
Traffic analysis, as presented in \cite{Hullar2014}, has been leveraged by network operators to effectively manage and monitor their networks. These operators often aim to identify the applications generating traffic within their networks and optimize their performance. State-of-the-art packet-based traffic classification methods typically rely on either expensive payload inspection of multiple packets per flow or basic flow statistics that do not consider packet content.

Additionally, traffic monitoring plays a significant role in identifying hidden network security risks that can disrupt normal network operation. Therefore, to enhance network security, monitoring abnormal network traffic becomes imperative. However, traditional network traffic anomaly monitoring systems often suffer from low accuracy and long monitoring times. To address this challenge, \cite{Huang2023} proposed a data mining-based network traffic anomaly monitoring system. In another approach, \cite{Bai2022} tackled the issue of monitoring and identifying abnormal network traffic by employing mutual information-based feature selection and comparing various mathematical models for classification.

The breadth of these studies illustrates the wide-ranging domain covered by network traffic monitoring and underscores its significance in network operations and security.

\subsection{Website Fingerprinting}
\label{website_fingerprinting}
Website Fingerprinting (WF) is a form of traffic analysis attack that enables eavesdroppers to discern a user's web activity, even when privacy tools like proxies, VPNs, or Tor are employed. Despite previous defense methods against WF attacks, newer attacks utilizing more sophisticated classifiers have rendered many of these defenses ineffective. Furthermore, the few remaining defenses considered effective tend to be inefficient, resulting in poor user experience and imposing significant overhead on servers \cite{Shan2021}. Notably, WF attacks can unveil the websites to which users are connected by analyzing encrypted traffic between users and anonymous network portals. Recent research has demonstrated the feasibility of launching WF attacks on anonymous Tor networks, even with a limited number of samples \cite{Wang2013}.

The threat posed by website fingerprinting has garnered substantial attention in recent years, particularly due to its implications for user privacy in the face of Internet censorship, surveillance, and tracking. Consequently, multiple studies have been conducted to analyze and mitigate the risks associated with trace-based attacks on internet pages. This section provides a comprehensive overview of these studies, emphasizing their contributions and limitations. The body of similar work in the field of page trace-based attacks offers valuable insights into the latest techniques and challenges, crucial for researchers and practitioners aiming to develop effective defenses that enhance user privacy and security in the online realm.

Anonymity stands as one of the most critical security concerns on the Internet, and despite users' incomplete awareness of the risks involved, numerous attacks have been devised to exploit these vulnerabilities. With unlimited access to the Internet, users can easily engage in illicit and socially harmful activities, such as those found on the dark web. \cite{Akshobhya2021} propose various methods for classifying website visits at different levels, shedding light on the anonymity-related challenges.

Building on the importance of page traces based on TCP/IP headers, as highlighted in \cite{Yan2018}, previous studies have primarily focused on a limited set of features. However, this study conducts a comprehensive analysis of communication features, identifying previously unaddressed attributes that can be employed to classify network traffic. The findings contribute to understanding the potential impact of information extraction capabilities in carrying out an attack.

In summary, the prominence of website fingerprinting as a traffic analysis attack calls for extensive research efforts to develop effective defenses against it. By delving into the latest techniques, challenges, and insights presented in these studies, researchers and practitioners can work towards enhancing user privacy and security in an ever-evolving online landscape.

\section{Digital Fingerprints}
\label{digital_fingerprints}
Every day, as we engage in our physical endeavors, we come into contact with various objects that we touch. With each touch, we unwittingly leave behind a unique set of fingerprints, tangible evidence of our interaction with these objects. If a meticulous inspection were conducted on all the physical entities we touch within a given space, a wealth of information could be extracted, illuminating our actions and behaviors. The physical evidence at our disposal encompasses which objects were touched, the manner in which they were handled, their condition (whether pristine or damaged), and the potential implications of these interactions. Such revelations provide invaluable insight into the motives behind these actions. Even in cases where the precise individuals responsible for touching these objects are unknown, by comparing the fingerprints against an extensive repository of samples, we can successfully identify the person or, at the very least, narrow down the pool of likely candidates based on the sequence of actions and the behavioral patterns associated with them.

Parallel to our tangible activities, our virtual presence also leaves behind traces in the digital realm. These traces, commonly referred to as digital fingerprints, serve as distinct markers that identify both us and our devices in any corner of the vast Internet landscape. Therefore, digital fingerprints can be understood as remnants of information that we unwittingly leave behind while navigating the online sphere. These traces have the potential to unveil users' activities on the Internet, encompassing their browsing history, utilized applications, and even the devices employed during these engagements. In essence, digital fingerprints function as unique identifiers that enable the identification and monitoring of individuals' online behaviors.

The right to privacy is an essential entitlement that every individual possesses and should be able to exercise. In the digital age we inhabit today, the preservation of privacy rights has gained paramount significance, particularly as individuals share an increasing amount of personal information on the Internet. This right encompasses the authority to retain control over one's personal data and to safeguard their activities and personal information from prying eyes. An individual can enjoy privacy on the Internet when there is an insurmountable barrier between their identity and the activities they undertake. During web browsing, for instance, a user's identity is associated with their IP address, while the web pages they explore serve as a testament to their online activities.

\begin{figure*}
  \includegraphics[width=\textwidth]{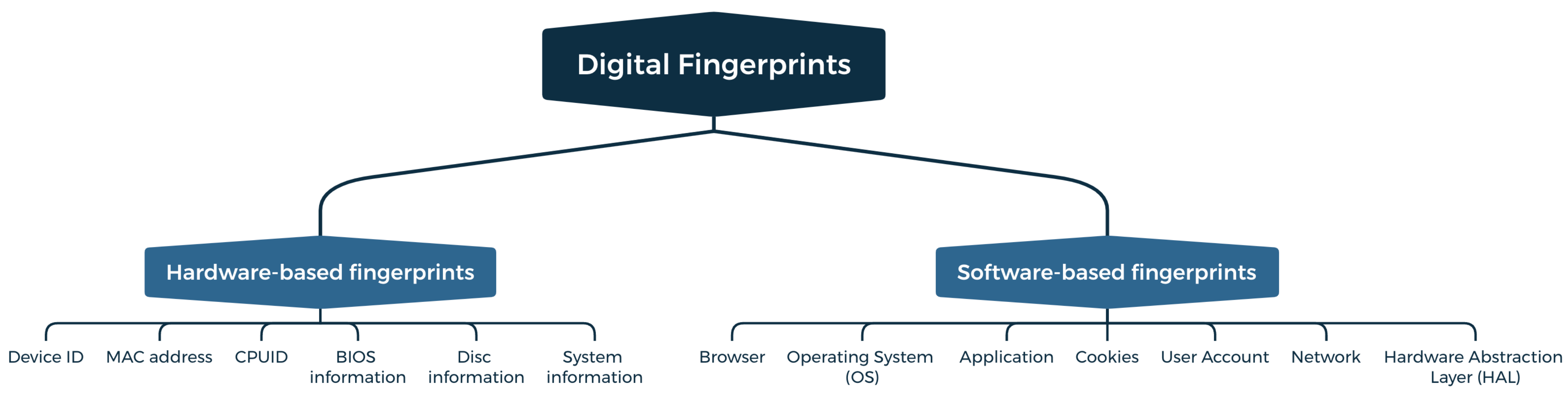}
\caption{Topology of various digital fingerprints}
\label{fig:digital_fingerprints}
\end{figure*}

The digital fingerprints left behind by users, which can readily lead to their identification, can be classified into two distinct categories: hardware-based traces and software-based traces, as illustrated in Fig. \ref{fig:digital_fingerprints}.

Hardware-based traces can be further sub categorized as follows:

\begin{enumerate}
    \item  Device ID: This refers to a distinctive identifier that is embedded within the device's hardware, often in the form of a serial number. It serves as a unique marker to differentiate one device from another \cite{patil2020hash}.
    \item MAC address: Every network card in a device is assigned a unique identifier known as a Media Access Control (MAC) address. This address is employed to identify and distinguish the device within a network environment \cite{takasu2015survey}.
    \item CPUID: The CPUID serves as a distinct identifier associated with the device's processor. It aids in the identification of the hardware, providing valuable information about the specific processor employed within the device \cite{NakiblySY15,laor2022drawnapart}.
    \item BIOS Information: This encompasses pertinent data pertaining to the device's Basic Input/Output System (BIOS). It includes details such as the manufacturer, version, and release date, which can be used to identify and characterize the device \cite{zhang2021digital}.
    \item Disk Information: This pertains to data relating to the device's storage disk, encompassing details such as the model and serial number. Such information aids in distinguishing one disk from another \cite{kumar2023device}.
    \item System Information: This category comprises data concerning the device's hardware configuration, offering insights into aspects such as the type and capacity of memory, processor specifications, and the model of the motherboard utilized \cite{bai2022passive}.
\end{enumerate}

By examining these hardware-based traces, one can gain valuable insights into the device itself, facilitating its identification and differentiation from other devices.

In contrast, software-based digital footprints can be categorized as follows \cite{Matos2021,Rizzo2021,Durey2022}:

\begin{enumerate}
 \item  Browser Fingerprints: These encompass a collection of data that serves as a unique identifier, derived from the characteristics and preferences of the web browser being utilized. Such information may include details like the browser version, installed plugins, or the availability of specific fonts. Extensive studies have been conducted on these browser traces, as documented in the references \cite{laperdrix2020browser,iqbal2021fingerprinting}. 

 \item Operating System (OS) Fingerprints: This pertains to information regarding the device's operating system, encompassing details such as its version and architecture. These attributes play a crucial role in identifying and understanding the software environment in which the device operates \cite{anderson2017fingerprinting}.

 \item Application Fingerprints: This category entails information concerning the software applications installed on the device. It encompasses details such as the version numbers of the applications and the specific software packages that have been installed \cite{ahmed2018statistical}.

 \item  Cookies: Cookies are sets of data stored by the web browser, which are associated with a user's browsing history. They can be employed to track the user's activities across multiple websites, providing insights into their online behavior and preferences \cite{boda2012user,fifield2015fingerprinting}.

 \item  User Account Fingerprints: This encompasses information pertaining to the user account associated with the device, including the username and user ID. Such traces contribute to the identification and association of specific user profiles \cite{nikiforakis2013cookieless}.

 \item  Network Fingerprints: Network traces encompass data relating to the network infrastructure to which the device is connected. This includes information such as the device's IP address, port configurations, and communication data. Such traces offer insights into the network context and connectivity patterns \cite{abdelnur2008advanced}.

 \item  Hardware Abstraction Layer (HAL) Fingerprints: This category encompasses information about the hardware abstraction layer of the device. The HAL serves as a vital component that bridges the hardware and software layers of the device's operating system. Traces at this layer provide valuable insights into the underlying hardware and its interaction with the software environment \cite{franklin2006passive,mirza2011fingerprinting}.
\end{enumerate}

By analyzing these software-based digital fingerprints, one can acquire comprehensive information about the browser, operating system, applications, user account, network context, and hardware abstraction layer of a device, thereby enhancing the ability to identify and understand the digital activities associated with it.

Web fingerprint based attacks are a class of attacks employed to discern the identity of users by exploiting their distinctive digital footprints. These attacks involve meticulously monitoring a user's network traffic and scrutinizing the intricate patterns embedded within their online activities. Through meticulous analysis of these patterns, assailants can successfully deduce the specific Internet pages being accessed by the user. Such attacks fall within the purview of traffic analysis, granting local network eavesdroppers the ability to clandestinely identify the webpages that a user visits. Although the presence of local network eavesdroppers remains imperceptible, their impact on user privacy is grave.

The modus operandi of these attacks entails the attacker initially acquiring knowledge of the user's browsing identity and subsequently amassing observable client traffic in the form of sequential packets. By employing machine learning classification techniques to analyze the client's packet sequence, the eavesdropper can accurately discern the precise webpage being browsed by the user. Consequently, the eavesdropper successfully re-establishes the link between the client's identity and their browsing activities, thereby circumventing the intended separation of these two realms and exacerbating the infringement on the client's privacy.

\begin{figure*}
  \includegraphics[width=\textwidth]{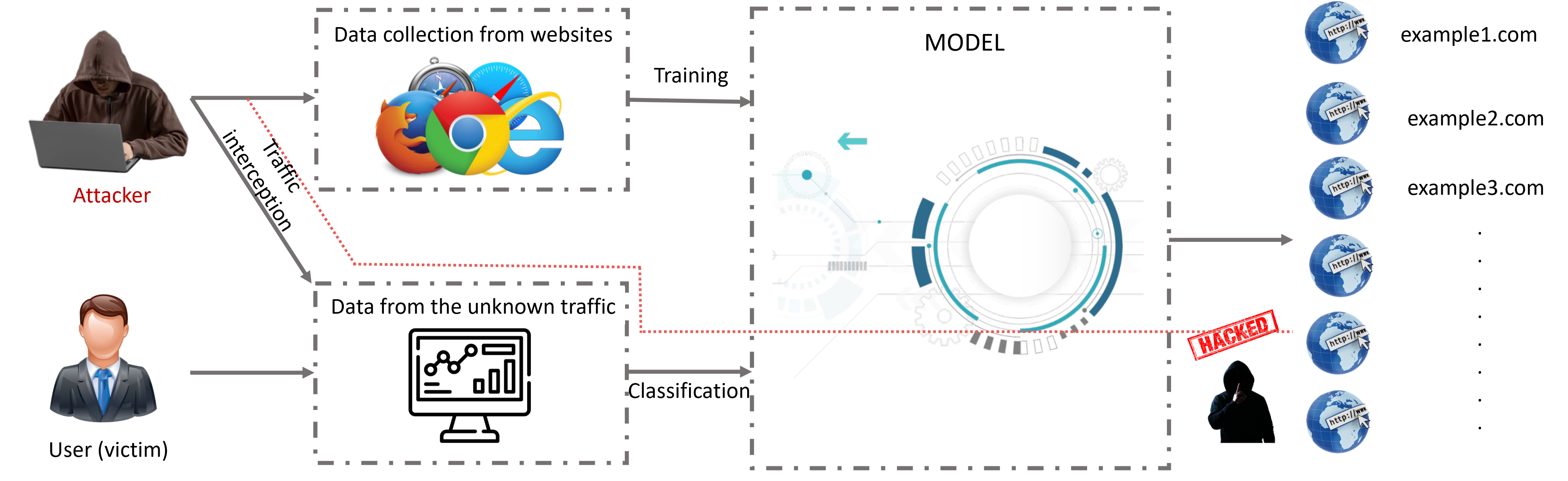}
\caption{Attack development model}
\label{fig:threat_model}
\end{figure*}

Attacks predicated on web page traces leverage an array of sophisticated techniques, including statistical analysis, machine learning algorithms, and pattern recognition methodologies. These multifaceted attacks excel in identifying web pages characterized by distinct activity patterns or belonging to specific domains of use.

The realm of attacks based on page traces presents an inherent dichotomy between security and privacy. On one hand, the ability of attackers to discern a user's communication flow by leveraging readily accessible internet data during web browsing, without necessitating the decryption of page packets, engenders profound privacy concerns. Conversely, such applications offer the potential to identify potentially malicious or suspicious behavior through the astute analysis of traffic patterns within a network. This capability proves invaluable in identifying and thwarting attacks while detecting anomalous activities indicative of security breaches. Moreover, web page traces can aid in the identification and prioritization of crucial network traffic, such as critical applications or high-priority communications. This strategic prioritization ensures that vital traffic remains unhindered by less consequential data or deliberate attacks, thereby upholding the integrity and continuity of essential network operations.

\section{Experimental Setup}
\label{experimental_setup}

\subsection{Threat Model}
\label{threat_model}
Within the context of this case study, we adopt a threat model that revolves around a passive local adversary who possesses unrestricted access to the targeted network and possesses the capability to monitor the fingerprints of users' Internet communications within that network. However, it's important to note that this attacker does not engage in activities aimed at influencing the traffic flow, such as introducing delays or modifying packets.  Instead, his role is solely focused on monitoring the fingerprints of users' Internet communications. Consequently, this attacker is referred to as a passive attacker. 

The primary objective of the passive attacker is two-fold. Firstly, they aim to differentiate the traffic at a binary level, determining whether the observed fingerprints belong to the browsing activity of any of the pages from a predefined list. This allows them to identify whether a user has visited a website of interest. Secondly, the attacker seeks to ascertain precisely which pages from the predefined list were visited. They achieve this by employing multi-class classification techniques, which enable them to classify the observed traffic into specific website categories.

To execute the attack, the attacker must initially collect data from the network by visiting a set of selected pages. This data serves as a training set for their classifier. Given the sheer vastness of the internet,  it is impractical to collect data from visits to every single website. Therefore, the monitoring is constrained to a predefined list of websites, termed as the list of monitored pages. Any pages visited outside this predefined list fall into the unmonitored category. Once the classifier has been trained and demonstrates a significant level of accuracy, the attack can be initiated. 

During the attack, the passive attacker intercepts the traffic flow of users within the network. They extract communication features similar to those employed during the training phase of the classifier. These features capture various aspects of the network traffic, allowing the attacker to identify patterns and unique characteristics associated with specific websites. Subsequently, the classification process is performed on the extracted features, enabling the attacker to identify the specific pages visited by the users.

By adopting this threat model and outlining the step-by-step process of the attack, we gain insights into the attacker's capabilities, limitations, and objectives. This provides a clear framework for understanding the risks associated with website fingerprinting attacks and underscores the need for effective countermeasures to protect user privacy in scenarios where passive local adversaries are present.

A comprehensive depiction of the steps involved in implementing this threat model is presented in Fig. \ref{fig:threat_model}.

\subsection{Implementation Steps}
\label{impementation_steps}


The implementation process of this case study involved a series of well-defined steps, outlined as follows:
\begin{itemize}
    \item Data Collection: Data or exchanged packets of network traffic were gathered utilizing dedicated tools specifically designed for this purpose or through custom scripts developed for packet monitoring.
    \item Packet Feature Identification and Extraction: Relevant packet features from the network traffic data were identified and extracted. These features encompassed properties such as packet size, time intervals between packets, direction, protocol, and other pertinent attributes.
    \item Data Processing: The collected data was processed by associating each packet in the dataset with the corresponding visited web page. This entailed tagging the data for each network traffic trace. Additionally, processing steps encompassed the removal of duplicate data, handling missing values, encoding categorical features, and normalizing and scaling features to ensure they were on a consistent scale. Feature engineering techniques were also applied to select and transform features, aiming to enhance the performance of the machine learning model.
    \item Machine Learning Algorithm Selection: An appropriate machine learning algorithm was chosen for the classification task. Several options were considered, including decision trees, random forests, support vector machines, and deep learning algorithms. The choice of algorithm relied on the characteristics of the data and the desired performance metrics.
    \item Model Training: The selected machine learning model was trained using the processed data. The dataset was split into training, validation, and test data. The training data was utilized to train the model, with the model being fitted to the training data to learn the relationship between the input features and the output classification.
    \item Performance Evaluation: The performance of the trained model was evaluated using the validation dataset. Appropriate performance metrics such as accuracy, precision, recall, and F1-score were employed to assess the model's effectiveness.
    \item Model Testing: The trained model was tested using the test data to verify its accuracy and generalization capabilities.
    \item Model Tuning: In the event of unsatisfactory performance, the model was fine-tuned by adjusting hyperparameters or exploring alternative algorithms that may be better suited to the task.
    \item Model Deployment: The trained model was utilized to classify new network traffic data, providing real-time identification of web page visits.
\end{itemize}

\subsubsection{Network traffic generation and monitoring}
\label{traffic_generation}

In order to undertake the development of this case study, several meticulous steps were taken to prepare the environment, establish the necessary conditions, and acquire the essential data for its execution.

Foremost, the imperative consideration was the monitoring of network traffic and the collection of authentic data. Respecting the laws safeguarding citizen privacy, it would be unethical to monitor the network traffic to which individuals have access without their knowledge and consent. Consequently, such monitoring, given its intricate nature, was deemed unfavorable. To circumvent this issue, an approach involving the generation of artificial traffic and simultaneous monitoring as a test scenario was adopted.

As elucidated in Section \ref{related_work}, diverse tools for network traffic monitoring exist, providing potential means to carry out the monitoring. However, preliminary trials revealed that employing these tools within a non-isolated network would not only capture traffic from our own devices but also from other individuals' devices. Although it might have been possible to filter and extract solely the network packets exchanged by our devices from the collected data, this approach would still linger on the ethical edge due to the possession of unfiltered, untargeted data from other network users. Consequently, a definitive decision was made to monitor the traffic programmatically through a script, which will be presented subsequently, explicitly confining the monitoring solely to the communications generated by us. As a result, during the monitoring process, only packets exchanged by personal devices were observed, recorded, and analyzed, distinctly differentiating them from the network activity of other users.

\begin{table*}[ht]
\centering
\begin{tabular}{p{0.08\linewidth}p{1.5cm}p{1.5cm}p{1.8cm}p{1.8cm}p{1.2cm}p{1.2cm}p{1.5cm}}
  \hline
Flow duration & Fwd packets & Bwd packets & Length of fwd packets & Length of bwd packets & Flow bytes & Flow packets & Average packet size \\
\hline	
1.911063 & 109 & 109 & 7953	& 0	& 4161.56 & 57.04	& 72.96 \\
0.138093 & 7 &7 &	957	&0	&6930.11	&50.69	&136.71	 \\
	1.869037 &	156	& 156	& 206294	& 0	& 110374.49	&83.47 & 1322.40	 \\
	0.096231	& 5&	5&	2397	&0	&24908.81&	51.96&479.4	 \\
0.131105& 	6	&6	&938&	0	&7154.57	&45.76	&156.33	 \\

... & ... & ... & ... & ... & ... & ... & ...\\
\hline
\end{tabular}
\caption{Example of dataset structure}
\label{tbl:dataset}
\end{table*}

\subsubsection{Data Collection}
\label{data_collection}

In order to structure the data collection process and the data itself, we have taken as a reference and comparative point a public dataset which has been widely used in a considerable number of research papers \cite{Kaggle_Kameri1,Kaggle_Kameri2,Kaggle_Kameri3}.

The dataset at hand was meticulously gathered within the network infrastructure of Universidad Del Cauca, located in Popayán, Colombia \cite{KaggleDataSet}. It was compiled through thorough packet capture operations conducted at various time points, encompassing both morning and afternoon sessions, spanning a six-day period: April 26, 27, 28, as well as May 9, 11, and 15, all during the year 2017. This concerted effort yielded a substantial repository of 3,577,296 distinct instances, systematically organized in a Comma Separated Values (CSV) file format.

The dataset's content is characterized by a diverse array of 87 features. Each entry in this dataset encapsulates intricate details of an IP flow originating from different network devices. Noteworthy attributes of these entries include source and destination IP addresses, port identifiers, interarrival intervals, and notably, the layer 7 protocol, which denotes the specific application governing the data flow. This latter aspect assumes significance as it serves as the primary class identifier, revealing the underlying application being utilized. The dataset encompasses attributes of numeric, nominal, and date types, the latter capturing temporal details and chronological information through the Timestamp field. This dataset embodies a multifaceted collection of insights derived from network activities, poised to unveil behavioral patterns and facilitate strategic decision-making in the domains of network analysis and cybersecurity. 

This dataset stands out as a remarkable contribution within the realm of network traffic analysis due to its distinct focus on surpassing conventional boundaries. While numerous existing network traffic classification datasets predominantly concentrate on identifying broad categories of application types within IP flows, such as WWW, DNS, FTP, P2P, and Telnet, this dataset embarks on a pioneering trajectory by transcending these limitations. It takes a substantial leap forward by not only discerning the generic application types but, notably, by delving into the realm of specificity. By training sophisticated machine learning models on IP flow statistics, the dataset achieves the remarkable feat of detecting and distinguishing individual applications with unparalleled precision, encompassing prominent platforms like Facebook, YouTube, Instagram, and an array of 75 distinct applications. This progressive endeavor elevates the dataset to a unique echelon, enabling the exploration of finer nuances within network behaviors and offering a fertile ground for advancing predictive analytics in the context of network flow analysis.

Following a similar path in order to collect the data for the purpose of traffic monitoring, a series of preparatory steps were diligently followed. Given the need to generate the traffic ourselves, the scope was confined to a specific domain, wherein a curated list of Internet pages to be visited was defined. Specifically, the decision was made to monitor the traffic generated during the browsing of online media sites from a pre-established list. The sequential steps undertaken in this process are outlined below:

\begin{itemize}
    \item Selection of Media Sources: A comprehensive compilation of written or televised media outlets, as documented in Annex [reference to the annex], was meticulously curated.
    \item URL Verification: The URLs associated with the selected media sources were meticulously obtained and subjected to thorough accessibility checks to ensure their availability.
    \item Synchronization of Traffic Generation and Monitoring: A synchronized pair of functions were meticulously devised and implemented to facilitate the simultaneous generation of traffic while monitoring the ensuing network activity.
    \item Iterative Traffic Generation: The predetermined list of media sites, as determined through the preceding steps, was repeatedly browsed to generate the desired traffic, ensuring a comprehensive dataset.
\end{itemize}

By diligently following these steps, the essential groundwork was laid to enable the subsequent analysis and exploration of the gathered traffic data.

\subsubsection{Data Preprocessing}
\label{data_preprocessing}
The collected data, as elucidated earlier, is stored in the .pcapng format, necessitating conversion and processing for further analysis. Initially, it is imperative to classify the packets based on their distinct characteristics and group them into separate flows. Subsequently, the flow data must be transformed into an acceptable format for future modeling, such as .csv.

The aggregated packet data contained within the .pcapng files cannot be directly utilized in its current form; thus, a series of preprocessing steps are undertaken to prepare the data for subsequent analysis. The following steps outline the data processing procedure:

\begin{table}[b]
\centering
\begin{tabular}{llll}
  \hline
Time &  Packet number & Flow number\\
\hline	
\hline	
Day 1  &  2238327  &  3613 \\
Day 2  &  2113341  &  1403 \\
Day 3  &  2051950  &  1395 \\
Day 4  &  1442588  &  1145 \\
Day 5  &  624371  &  497 \\
Day 6  &  1186974  &  817 \\
Day 7  &  730878  &  491  \\
Day 8  &  1031972  &  1711  \\
Day 9  &  1856545  &  3542  \\
Day 10  &  885234  &  1565  \\
Day 11  &  3130620  &  5557  \\
\hline
\hline
Total &  17292800  &  21736  \\
\hline
\end{tabular}
\caption{Traffic data statistics}
\label{tbl:packet_number}
\end{table}

\begin{enumerate}
    \item  Extraction of Visit-specific Packets: The packets pertaining to each visit, as stored in the .pcapng format, are extracted, isolating the relevant packets associated with individual browsing sessions.
    \item  Classification and Flow Collection: The packets are classified based on their distinct characteristics and organized into separate flows, ensuring a coherent representation of the network communication patterns.
    \item Conversion to Excel Format: The extracted flows are converted into the Excel format, enabling efficient data manipulation and processing.
    \item Tagging of Streams: Each extracted flow is meticulously tagged according to the corresponding visited page, providing context and facilitating subsequent analysis.
    \item Saving Extracted Streams: The visit data, accompanied by the corresponding extracted flows, is persistently stored in .csv format, ensuring compatibility and ease of use for future analysis.
    \item Merging the Dataset: A new dataset is created, encompassing the structured representation of visits with their associated packet flows. This comprehensive dataset incorporates all the previously extracted flows, ensuring a cohesive and unified data structure.
\end{enumerate}

An exemplification of the final dataset structure resulting from the aforementioned processing steps is depicted in Table \ref{tbl:dataset}.

The data collected, as elucidated earlier, constitutes a closed world dataset, encompassing a restricted collection of network packets. This implies that the dataset exclusively comprises the targeted packets exchanged during browsing specific pages, while excluding any extraneous packets. Moreover, these packets are recorded within a defined time frame and under controlled circumstances, ensuring the dataset's integrity and specificity.

To augment the dataset, periodic re-browsing was conducted for all selected sites, facilitating the acquisition of additional test data over a span of two weeks subsequent to the initial data collection phase. Table \ref{tbl:packet_number} illustrates the number of packets and flows observed during each day of the monitoring period, providing an overview of the dataset's temporal distribution.

To facilitate the implementation of a two-layer classification, where data is first classified based on its relevance to the model and subsequently categorized, the dataset has been divided and labeled accordingly. Specifically, all newspaper pages associated with Albania or Kosovo have been marked as targeted pages, deemed of interest to the model. Conversely, the remaining pages have been designated as random pages. This partitioning strategy ensures that the dataset primarily focuses on a specific subset of pages. It should be noted that while these sites represent a portion of the online presence within the two regions, they are the sole ones labeled as such within the dataset. The partitioning based on targeting is presented numerically in Table \ref{tbl:datasetNdaras}, which will be leveraged for the implementation of binary classification.

\begin{table}[ht]
\centering
\begin{tabular}{ll}
    &   Flow number\\
\hline	
	
Targeted data  &  4699\\
Untargeted data  & 17026\\
\hline
\hline
Total  & 21725  \\

\end{tabular}
\caption{Separated data}
\label{tbl:datasetNdaras}
\end{table}

\subsubsection{Hyperparameter Selection}
\label{hyperparameter_selection}

To construct the machine learning models for our case study, it is crucial to establish initial models to fine-tune the values of hyperparameters, followed by building optimal models that leverage the most effective combinations of these hyperparameters, ensuring superior performance.

Cross-validated grid search optimization serves as a valuable technique for identifying the optimal hyperparameters for machine learning models. The steps involved in configuring the hyperparameters of the machine learning algorithms are outlined as follows:

\begin{itemize}
    \item Definition of Hyperparameters: The hyperparameters to be fine-tuned are carefully specified.
    \item Data Split: The dataset is divided into training and validation data. The training data is utilized for training the model, while the validation data is employed for evaluating its performance.
    \item Grid Definition: A grid, comprising all possible combinations of hyperparameters to be tuned, is constructed.
    \item Cross-Validation: Cross-validation is employed to assess the performance of each hyperparameter combination. For every combination within the grid, a model is trained using the training set, and its performance is evaluated through cross-validation. The average cross-validation score serves as the performance metric for the specific hyperparameter combination.
    \item Optimal Hyperparameters Selection: The hyperparameter combination yielding the best performance is selected.
    \item Model Evaluation: The model's performance is evaluated using the optimal hyperparameters on the validation data. Subsequently, the model is fitted using the optimal hyperparameters and the entire training dataset. Finally, its performance is assessed using the validation dataset.
\end{itemize}

By following these steps, we can effectively determine the most suitable hyperparameters for our machine learning models, leading to enhanced performance and accuracy.

\subsubsection{Model training and evaluation}
\label{model_training}

In the pursuit of developing a robust predictive framework for discerning the URLs associated with network flows, a comprehensive ensemble of diverse machine learning models was harnessed. The range of these models encompassed: (i) Decision Tree (DT), (ii) Random Forest (RF), (iii) Gradient Boosting Machine (GBM), (iv) Adaptive Boosting (AdaB), (v) Support Vector Machine (SVM), (vi) Naive Bayes (NB), and (vii) k-Nearest Neighbors (KNN). Each of these models was harnessed to traverse the intricate landscape of network flow data, aiming to unravel patterns and relationships that could be indicative of the pages being visited.

The utilization of these distinct models, characterized by their unique algorithmic underpinnings and predictive capabilities, fortified the exploration of the dataset's nuanced dynamics. Through a systematic process, the data was partitioned into training and validation subsets, allowing each model to undergo rigorous training while harnessing the former, followed by meticulous evaluation against the latter. This evaluation encompassed a gamut of metrics tailored to the nature of the classification task, encompassing precision, recall, F1-score, and accuracy. The collective outcomes of these model evaluations provided a comprehensive panorama of their respective performance in ascertaining the elusive pages linked with network flows. By amalgamating the discerning insights garnered from each model, this study strides toward establishing a comprehensive and informed foundation for the predictive analysis of network traffic.

To measure a model we use machine learning model evaluation metrics, as mentioned above. The choice of metric depends on the problem and the model. Considering that:

     \quad\textit{TP} is the number of true positive predictions,
    
     \quad\textit{FP} is the number of false positive predictions,
    
     \quad\textit{TN} is the number of true negative predictions,
    
     \quad\textit{FN} is the number of false negative predictions

below are listed some of the commonly used metrics for evaluating machine learning models with reference to \cite{Irsoy2013,Shervin2019}:
\begin{description}
\item [Accuracy] measures the percentage of correct model predictions by calculating the proportion of correctly classified cases out of the total number of cases in a data set. In other words, it measures how well the model is able to correctly predict the class labels of the input data. The accuracy can be calculated as follows:

\begin{equation}
\text{Accuracy} = \frac{\text{TP + TN}}{\text{TP + TN + FP + FN}}
\end{equation}

Accuracy is a useful metric when the classes in the dataset are balanced, meaning that there are approximately equal numbers of examples for each class.
\item [Precision] measures the fraction of true positive predictions among all positive predictions. So it measures how many of the cases that the model predicted as positive are actually positive. The precision can be calculated as follows:

\begin{equation}
\text{Precision} = \frac{\text{TP}}{\text{TP + FP}}
\end{equation}

Precision is a useful metric when the cost of false positive predictions is high, meaning that it is important to avoid positive predictions for negative cases. For example, in medical diagnosis, false positive predictions can lead to unnecessary treatment, which can be costly or even harmful to the patient. In such cases, high precision models are preferred over high draw models.
\item [Recall]: measures the fraction of true positive predictions among all actual positive cases. Recall is a commonly used machine learning evaluation metric that measures the fraction of true positive predictions among all positive instances in the data set. In other words, it measures how many of the actual positive cases in the data set are correctly identified by the model as positive. Recall can be calculated as follows:
\begin{equation}
\text{Recall} = \frac{\text{TP}}{\text{TP + FN}}
\end{equation}
Recall is a useful metric when the cost of false negative predictions is high, meaning that it is important to avoid negative prediction for positive cases. For example, in disease diagnosis, false negative predictions can delay or miss the diagnosis, which can be harmful to the patient. In such cases, high recall models are preferred over high accuracy models.
\item [F1 Score] is the harmonic mean of precision and recall with equal weights assigned to both metrics. It is calculated as in the following formula:
\begin{equation}
\text{F1 Score} = 2 * \frac{\text{Precision * Recall}}{\text{Precision + Recall}}
\end{equation}

The F1-score is a useful metric when accuracy and recall are important in a binary classification problem and there is no clear preference between these two metrics. It is also useful in situations where the distribution of classes is unbalanced and the performance of the model needs to be evaluated with respect to the positive and negative cases in the data set.
\end{description}

\section{Results}
\label{results}
Upon completion of model training using optimal parameters, an assessment was conducted to evaluate the predictive efficacy of both classification types. The resultant accuracy of the models' predictions was meticulously computed. Comprehensive performance outcomes of the model's proficiency are presented in the subsequent tables. Specifically, Table \ref{tbl:performanceB} showcases the outcomes derived from models trained for binary classification, whereas Table \ref{tbl:performanceMC} exhibits the findings of models trained for multi-class classification. Key metrics, including accuracy, precision, recall, and F1 rate, have been methodically documented for each machine learning model.

\begin{table}[ht]
\centering
\begin{tabular}{lllll}
  \hline
Model & Accuracy & Precision & Recall & F1 Score\\
\hline	
\hline
DT & 0.8002 & 0.5615 & 0.4327 & 0.4888\\
RF & 0.8354 & 0.7053 & 0.4369 & 0.5396\\
\rowcolor[HTML]{2E668F} \color{white} GBM &  \color{white}0.8363 & \color{white} 0.7271 &  \color{white}0.4139 &  \color{white}0.5275\\
AdaB & 0.7958 & 0.6232 & 0.1897 & 0.2909\\
SVM & 0.7852 & 0.7096 & 0.0458 & 0.0861\\
NB & 0.7675 & 0.1772 & 0.0145 & 0.0269\\
KNN & 0.7926 & 0.5587 & 0.2877 & 0.3799\\
\hline
\end{tabular}
\caption{Performance metrics values of binary classification models}
\label{tbl:performanceB}
\end{table}

In the realm of binary classification, the pinnacle of accuracy was reached by the Gradient Boosting Machines algorithm, boasting an impressive accuracy score of 0.8363. Complementing this accomplishment, the model exhibited noteworthy precision (0.7271), recall (0.4139), and F1 score (0.5275). Following suit, the Random Forest, Decision Tree, Adaptive Boost, and Support Vector Machines models demonstrated respectable performance. In contrast, the Naive Bayes algorithm yielded the least accurate results, with an accuracy of 0.7675, precision of 0.1772, recall of 0.0145, and F1 score of 0.0269.

Shifting our focus to multi-class classification, the Random Forest algorithm emerged triumphant, achieving a commendable accuracy rating of 0.6297. Alongside this feat, the model showcased commendable precision (0.6476), recall (0.6297), and F1 score (0.6316). In sequence, the Gradient Boosting Machines, Decision Tree, K-Nearest Neighbors, Adaptive Boost, Naive Bayes and Support Vector Machines models exhibited competitive performances. Conversely, the KNN algorithm struggled to deliver accurate results, attaining a meager accuracy of 0.4797, precision of 0.4759, recall of 0.4797, and F1 score of 0.4746.

The outcomes of the Adaptive Boost, Naive Bayes, and SVM models in multi-class classifications did not demonstrate a level of performance that could be deemed as consequential in the context of this study. Consequently, these results were not included in the result tables to provide a clear and focused presentation of the most impactful findings.

In the multi-class classification domain, the Adaptive Boost, Naive Bayes, and SVM models failed to attain the levels of accuracy and precision exhibited by the leading models, such as the Gradient Boosting Machines and Random Forest. Their performance, while not negligible, did not significantly contribute to the comprehensive insights provided by the study's primary outcomes.

Thus, in the interest of clarity and emphasizing the most influential model performances, the results, presented in Table \ref{tbl:performanceMC}, focuses on the more substantial outcomes achieved by the Decision Tree, Random Forest, Gradient Boosting Machines, and other noteworthy models. The models with comparatively less impact, namely Adaptive Boost, Naive Bayes, and SVM, were omitted to ensure a concise and informative presentation of the study's core findings.

\begin{table}[ht]
\centering
\begin{tabular}{lllll}
  \hline
Model & Accuracy & Precision & Recall & F1 Score \\
\hline
\hline
DT & 0.5510 & 0.6239 & 0.5510 & 0.5716\\
\rowcolor[HTML]{2E668F} \color{white} RF &  \color{white}0.6297 & \color{white} 0.6476 &  \color{white}0.6297 &  \color{white}0.6316\\
GBM & 0.6234 & 0.6216 & 0.6234 & 0.6180\\
KNN & 0.4797 & 0.4759 & 0.4797 & 0.4746\\

\hline
\end{tabular}
\caption{Performance metrics values of multi-class classification models}
\label{tbl:performanceMC}
\end{table}

In some binary classification models it can be seen that some metrics have shown much higher results than others. this is directly related to the way these metrics are calculated by referring to the Section \ref{model_training}. Therefore, comparison of the results from \cite{Kaggle_Kameri3} without its associated approach is not strait forward with  results presented in Table \ref{tbl:performanceMC}, even this leads to the same best model, with light difference in precesion and recall values. Considering the fact that the dataset used in our case study is significantly smaller and also contains a smaller number of features that were taken into consideration during model training, it cannot be compared directly with the work in \cite{Kaggle_Kameri3}.

These classifiers are not specifically designed to handle imbalanced classes, it might prioritize classifying instances into the majority class in order to maximize overall accuracy. The classifier may be correctly classifying the majority of web page flows, leading to a high overall accuracy. This is because the classifier is performing well on the dominant class, which contributes more to the accuracy calculation. However, when it comes to the rare class, the classifier might struggle due to the limited number of examples and the potential complexity of the class. As a result, the classifier may incorrectly classify some of theflow, leading to a low precision. In other words, false positives are driving down the precision for the rare class.


\section{Conclusions}
\label{conclusions}

The outcomes revealed within the Results section substantiate the noteworthy efficacy of the employed machine learning algorithms, particularly given the context of the dataset's modest scale, containing a mere 21,736 records. These results are indeed quite commendable, showcasing the models' adeptness at distilling intricate patterns from limited data.

In the domain of binary classification, the Gradient Boosting Machines algorithm demonstrated exceptional prowess, achieving accuracy score of 0.8363. This feat, where the model correctly predicted the target outcomes for approximately 83.63\% of instances, is particularly impressive given the dataset's size. The elevated precision (0.7271), which signifies the proportion of true positive predictions among all predicted positives, exemplifies the model's ability to accurately classify relevant instances. Additionally, the substantial recall (0.4139), indicating the proportion of true positive predictions among actual positives, illustrates the model's capacity to capture a significant portion of pertinent instances. The corresponding F1 score (0.5275), harmonizing precision and recall, further underscores the robustness of this model's predictive performance.

Transitioning to the realm of multi-class classification, the Random Forest algorithm's achievement of an accuracy rating of 62.97\% is particularly notable. In a dataset of this size, where classes could be sparsely represented, such an accuracy level signifies the model's ability to make well-informed predictions across multiple categories. This is corroborated by the commendable precision (0.6476), recall (0.6297), and F1 score (0.6316) achieved by the model, demonstrating its competence in effectively identifying and differentiating between diverse classes, even with the inherent challenges posed by a limited dataset.

In essence, the laudable performance of these machine learning models, despite the dataset's relatively modest scale, speaks volumes about their adaptability and ability to extract meaningful insights from constrained data. The results underscore the potential of these algorithms to generalize well, capturing intricate relationships and patterns even with limited samples. This successful endeavor not only validates the rigorous methodology employed but also accentuates the models' capacity to excel in scenarios where data availability is constrained.

On the other hand, while analyzing the results, the observed discrepancies in accuracy between the models—Decision Tree, Random Forest, GBM, and k-Nearest Neighbors (KNN)—versus the models Adaptive Boost, Support Vector Machine (SVM), and Naive Bayes—can be attributed to inherent differences in their underlying algorithmic methodologies and how they handle the complexity and nature of the network flow dataset.

The Decision Tree, Random Forest, GBM, and KNN models are known for their intrinsic ability to capture intricate relationships and patterns within the data. Decision Trees can discern hierarchical decision boundaries, while Random Forest and GBM are adept at aggregating multiple decision trees to collectively make informed predictions. KNN, on the other hand, leverages the proximity of data points to classify instances. In the context of network flows, where features might exhibit nonlinear and complex interdependencies, these models excel at capturing the nuanced behavior associated with different URLs. Their adaptability to diverse patterns enables them to effectively learn the intricate mappings between network flow attributes and visited URLs, resulting in comparatively higher accuracy.

Conversely, Adaptive Boost, SVM, and Naive Bayes may exhibit lower accuracy due to their unique characteristics and potential limitations in handling certain aspects of the dataset. Adaptive Boost, while a powerful ensemble technique, can struggle with noisy or mislabeled data, potentially compromising its performance on network flow data that could contain inherent noise. SVM, despite its strength in separating complex data, may face challenges when dealing with high-dimensional feature spaces commonly found in network flow datasets. Additionally, the kernel function choice and parameter tuning can significantly impact SVM's performance. Naive Bayes, while efficient and quick, relies on the assumption of feature independence, which might not hold in the context of network flows where attributes can exhibit intricate correlations.

In summary, the variations in accuracy between these model categories stem from their distinctive capabilities in handling the inherent complexities of network flow data. Decision Tree, Random Forest, GBM, and KNN models possess the versatility to capture intricate patterns, making them well-suited for this task, while Adaptive Boost, SVM, and Naive Bayes might face challenges related to noise, feature space dimensionality, and assumptions that can influence their predictive accuracy in this specific context.

\section*{Declarations}

\noindent \textbf{Research data policy and data availability} We have collected our test data as described in Section \ref{data_collection} with Wireshark software tool, stored it in the .pcapng format, and after the data preprocessing as a .csv file which is publicly available on \cite{dataset}. \\

\noindent \textbf{Competing Interests} All authors declare that they have no competing interest in the subject matter or materials discussed in this manuscript. \\

\noindent \textbf{Compliance with Ethical Standards} All authors declare that they comply with the ethical principles of the journal.


\bibliographystyle{ieeetr}
\bibliography{paper}

\begin{thebibliography}{10}

\bibitem{HTTPS}
Google, ``{Google Transparency Report}.''
  \url{https://transparencyreport.google.com/https/overview?hl=en/}, 2023.
\newblock [Online; accessed 19-July-2023].

\bibitem{kovalchuk2021dark}
O.~Kovalchuk, M.~Masonkova, and S.~Banakh, ``The dark web worldwide 2020:
  Anonymous vs safety,'' in {\em 2021 11th International Conference on Advanced
  Computer Information Technologies (ACIT)}, pp.~526--530, 2021.

\bibitem{liu2023survey}
P.~Liu, L.~He, and Z.~Li, ``A survey on deep learning for website
  fingerprinting attacks and defenses,'' {\em IEEE Access}, vol.~11,
  pp.~26033--26047, 2023.

\bibitem{Chaabane2010}
C.~Abdelberi, P.~Manils, and M.~A. K{\^a}afar, ``Digging into anonymous
  traffic: A deep analysis of the tor anonymizing network,'' {\em 2010 Fourth
  International Conference on Network and System Security}, pp.~167--174, 2010.

\bibitem{Cadena2020}
W.~De~la Cadena, A.~Mitseva, J.~Hiller, J.~Pennekamp, S.~Reuter, J.~Filter,
  T.~Engel, K.~Wehrle, and A.~Panchenko, ``Trafficsliver: Fighting website
  fingerprinting attacks with traffic splitting,'' in {\em Proceedings of the
  2020 ACM SIGSAC Conference on Computer and Communications Security}, CCS '20,
  (New York, NY, USA), p.~1971–1985, Association for Computing Machinery,
  2020.

\bibitem{MOUSTAFA2019}
N.~Moustafa, J.~Hu, and J.~Slay, ``A holistic review of network anomaly
  detection systems: A comprehensive survey,'' {\em Journal of Network and
  Computer Applications}, vol.~128, pp.~33--55, 2019.

\bibitem{GIBERT2020}
D.~Gibert, C.~Mateu, and J.~Planes, ``The rise of machine learning for
  detection and classification of malware: Research developments, trends and
  challenges,'' {\em Journal of Network and Computer Applications}, vol.~153,
  p.~102526, 2020.

\bibitem{Arbena}
A.~Musa, K.~Vishi, and B.~Rexha, ``Attack analysis of face recognition
  authentication systems using fast gradient sign method,'' {\em Applied
  Artificial Intelligence}, vol.~35, no.~15, pp.~1346--1360, 2021.

\bibitem{Blerim}
B.~Rexha, G.~Shala, and V.~Xhafa, ``Increasing trustworthiness of face
  authentication in mobile devices by modeling gesture behavior and location
  using neural networks,'' {\em Future Internet}, vol.~10, no.~2, pp.~1--17,
  2018.

\bibitem{Rrezearta}
R.~Thaqi, K.~Vishi, and B.~Rexha, ``Enhancing burp suite with machine learning
  extension for vulnerability assessment of web applications,'' {\em Journal of
  Applied Security Research}, pp.~1--19, 2022.

\bibitem{Wireshark}
R.~Sharpe, E.~Warnicke, and U.~Lamping, ``{Wireshark User’s Guide}.''
  \url{https://www.wireshark.org/docs/wsug_html/}, 2023.
\newblock [Online; accessed 19-July-2023].

\bibitem{Siswanto2019}
A.~Siswanto, A.~Syukur, E.~A. Kadir, and Suratin, ``Network traffic monitoring
  and analysis using packet sniffer,'' in {\em 2019 International Conference on
  Advanced Communication Technologies and Networking (CommNet)}, pp.~1--4,
  2019.

\bibitem{Finamore2011}
A.~Finamore, M.~Mellia, M.~Meo, M.~M. Munafo, P.~D. Torino, and D.~Rossi,
  ``Experiences of internet traffic monitoring with tstat,'' {\em IEEE
  Network}, vol.~25, no.~3, pp.~8--14, 2011.

\bibitem{Dorado2012}
J.~L. Garcia-Dorado, A.~Finamore, M.~Mellia, M.~Meo, and M.~Munafo,
  ``Characterization of isp traffic: Trends, user habits, and access technology
  impact,'' {\em IEEE Transactions on Network and Service Management}, vol.~9,
  no.~2, pp.~142--155, 2012.

\bibitem{Hullar2014}
B.~Hullár, S.~Laki, and A.~György, ``Efficient methods for early protocol
  identification,'' {\em IEEE Journal on Selected Areas in Communications},
  vol.~32, no.~10, pp.~1907--1918, 2014.

\bibitem{Huang2023}
Y.~Huang and L.~Huang, ``Design of network traffic anomaly monitoring system
  based on data mining,'' in {\em Advanced Hybrid Information Processing}
  (W.~Fu and L.~Yun, eds.), (Cham), pp.~549--563, Springer Nature Switzerland,
  2023.

\bibitem{Bai2022}
B.~Bai, ``Monitoring and identification of abnormal network traffic by
  different mathematical models,'' {\em Journal of Cyber Security and
  Mobility}, 12 2022.

\bibitem{Shan2021}
S.~Shan, A.~N. Bhagoji, H.~Zheng, and B.~Y. Zhao, ``A real-time defense against
  website fingerprinting attacks,'' {\em CoRR}, vol.~abs/2102.04291, 2021.

\bibitem{Wang2013}
T.~Wang and I.~Goldberg, ``Improved website fingerprinting on tor,'' in {\em
  Proceedings of the 12th ACM Workshop on Workshop on Privacy in the Electronic
  Society}, WPES '13, (New York, NY, USA), p.~201–212, Association for
  Computing Machinery, 2013.

\bibitem{Akshobhya2021}
A.~K~M, ``Machine learning for anonymous traffic detection and
  classification,'' in {\em 2021 11th International Conference on Cloud
  Computing, Data Science \& Engineering (Confluence)}, pp.~942--947, 2021.

\bibitem{Yan2018}
J.~Yan and J.~Kaur, ``Feature selection for website fingerprinting,'' {\em
  Proceedings on Privacy Enhancing Technologies}, vol.~2018, pp.~200--219, 10
  2018.

\bibitem{patil2020hash}
R.~Y. Patil and S.~R. Devane, ``Hash tree-based device fingerprinting technique
  for network forensic investigation,'' in {\em Advances in Electrical and
  Computer Technologies: Select Proceedings of ICAECT 2019}, pp.~201--209,
  Springer, 2020.

\bibitem{takasu2015survey}
K.~Takasu, T.~Saito, T.~Yamada, and T.~Ishikawa, ``A survey of hardware
  features in modern browsers: 2015 edition,'' in {\em 2015 9th International
  Conference on Innovative Mobile and Internet Services in Ubiquitous
  Computing}, pp.~520--524, IEEE, 2015.

\bibitem{NakiblySY15}
G.~Nakibly, G.~Shelef, and S.~Yudilevich, ``Hardware fingerprinting using
  {HTML5},'' {\em CoRR}, vol.~abs/1503.01408, 2015.

\bibitem{laor2022drawnapart}
T.~Laor, N.~Mehanna, A.~Durey, V.~Dyadyuk, P.~Laperdrix, C.~Maurice, Y.~Oren,
  R.~Rouvoy, W.~Rudametkin, and Y.~Yarom, ``Drawnapart: A device identification
  technique based on remote gpu fingerprinting,'' {\em arXiv preprint
  arXiv:2201.09956}, 2022.

\bibitem{zhang2021digital}
Y.~J. Zhang, L.~Zhang, F.~Yang, and S.~Gu, ``Digital fingerprint extraction
  method of iot devices based on cryptography,'' in {\em Proceedings of the
  2021 9th International Conference on Information Technology: IoT and Smart
  City}, pp.~288--291, 2021.

\bibitem{kumar2023device}
V.~Kumar and K.~Paul, ``Device fingerprinting for cyber-physical systems: A
  survey,'' {\em ACM Computing Surveys}, vol.~55, no.~14s, pp.~1--41, 2023.

\bibitem{bai2022passive}
S.~Bai, H.~Kim, and J.~Rexford, ``Passive os fingerprinting on commodity
  switches,'' in {\em 2022 IEEE 8th International Conference on Network
  Softwarization (NetSoft)}, pp.~264--268, IEEE, 2022.

\bibitem{Matos2021}
G.~Matos and E.~Feitosa, ``Identificando indicadores de browser fingerprinting
  em páginas web,'' in {\em Anais do XX Simpósio Brasileiro em Segurança da
  Informação e de Sistemas Computacionais}, (Porto Alegre, RS, Brasil),
  pp.~478--483, SBC, 2020.

\bibitem{Rizzo2021}
V.~Rizzo, S.~Traverso, and M.~Mellia, ``{Unveiling Web Fingerprinting in the
  Wild Via Code Mining and Machine Learning},'' {\em Proceedings on Privacy
  Enhancing Technologies}, vol.~2021, no.~1, pp.~43--63, 2021.

\bibitem{Durey2022}
A.~Durey, {\em Leveraging browser fingerprinting to strengthen web
  authentication}.
\newblock PhD thesis, Université de Lille, 01 2022.

\bibitem{laperdrix2020browser}
P.~Laperdrix, N.~Bielova, B.~Baudry, and G.~Avoine, ``Browser fingerprinting: A
  survey,'' {\em ACM Transactions on the Web (TWEB)}, vol.~14, no.~2,
  pp.~1--33, 2020.

\bibitem{iqbal2021fingerprinting}
U.~Iqbal, S.~Englehardt, and Z.~Shafiq, ``Fingerprinting the fingerprinters:
  Learning to detect browser fingerprinting behaviors,'' in {\em 2021 IEEE
  Symposium on Security and Privacy (SP)}, pp.~1143--1161, IEEE, 2021.

\bibitem{anderson2017fingerprinting}
B.~Anderson and D.~McGrew, ``Os fingerprinting: New techniques and a study of
  information gain and obfuscation,'' in {\em 2017 IEEE Conference on
  Communications and Network Security (CNS)}, pp.~1--9, IEEE, 2017.

\bibitem{ahmed2018statistical}
M.~E. Ahmed, S.~Ullah, and H.~Kim, ``Statistical application fingerprinting for
  ddos attack mitigation,'' {\em IEEE Transactions on Information Forensics and
  Security}, vol.~14, no.~6, pp.~1471--1484, 2018.

\bibitem{boda2012user}
K.~Boda, {\'A}.~M. F{\"o}ldes, G.~G. Guly{\'a}s, and S.~Imre, ``User tracking
  on the web via cross-browser fingerprinting,'' in {\em Information Security
  Technology for Applications: 16th Nordic Conference on Secure IT Systems,
  NordSec 2011, Tallinn, Estonia, October 26-28, 2011, Revised Selected Papers
  16}, pp.~31--46, Springer, 2012.

\bibitem{fifield2015fingerprinting}
D.~Fifield and S.~Egelman, ``Fingerprinting web users through font metrics,''
  in {\em Financial Cryptography and Data Security: 19th International
  Conference, FC 2015, San Juan, Puerto Rico, January 26-30, 2015, Revised
  Selected Papers 19}, pp.~107--124, Springer, 2015.

\bibitem{nikiforakis2013cookieless}
N.~Nikiforakis, A.~Kapravelos, W.~Joosen, C.~Kruegel, F.~Piessens, and
  G.~Vigna, ``Cookieless monster: Exploring the ecosystem of web-based device
  fingerprinting,'' in {\em 2013 IEEE Symposium on Security and Privacy},
  pp.~541--555, IEEE, 2013.

\bibitem{abdelnur2008advanced}
H.~J. Abdelnur, R.~State, and O.~Festor, ``Advanced network fingerprinting,''
  in {\em Recent Advances in Intrusion Detection: 11th International Symposium,
  RAID 2008, Cambridge, MA, USA, September 15-17, 2008. Proceedings 11},
  pp.~372--389, Springer, 2008.

\bibitem{franklin2006passive}
J.~Franklin, D.~McCoy, P.~Tabriz, V.~Neagoe, J.~V. Randwyk, and D.~Sicker,
  ``Passive data link layer 802.11 wireless device driver fingerprinting.,'' in
  {\em USENIX Security Symposium}, vol.~3, pp.~16--89, 2006.

\bibitem{mirza2011fingerprinting}
M.~Mirza, P.~Barford, X.~Zhu, S.~Banerjee, and M.~Blodgett, ``Fingerprinting
  802.11 rate adaption algorithms,'' in {\em 2011 Proceedings IEEE INFOCOM},
  pp.~1161--1169, IEEE, 2011.

\bibitem{Kaggle_Kameri1}
J.~S. Rojas, {\'A}.~R. Gall{\'o}n, and J.~C. Corrales, ``Personalized service
  degradation policies on ott applications based on the consumption behavior of
  users,'' in {\em Computational Science and Its Applications -- ICCSA 2018}
  (O.~Gervasi, B.~Murgante, S.~Misra, E.~Stankova, C.~M. Torre, A.~M.~A. Rocha,
  D.~Taniar, B.~O. Apduhan, E.~Tarantino, and Y.~Ryu, eds.), (Cham),
  pp.~543--557, Springer International Publishing, 2018.

\bibitem{Kaggle_Kameri2}
J.~S. Rojas, {\'A}.~Rend{\'o}n, and J.~C. Corrales, ``Consumption behavior
  analysis of over the top services: Incremental learning or traditional
  methods?,'' {\em IEEE Access}, vol.~7, pp.~136581--136591, 2019.

\bibitem{Kaggle_Kameri3}
J.~S. Rojas, A.~Pekar, {\'a}.~Rend{\'o}n, and J.~C. Corrales, ``Smart user
  consumption profiling: Incremental learning-based ott service degradation,''
  {\em IEEE Access}, vol.~8, pp.~207426--207442, 2020.

\bibitem{KaggleDataSet}
J.~S. Rojas, ``{Kaggle Dataset: IP Network Traffic Flows Labeled with 75
  Apps}.''
  \url{https://www.kaggle.com/datasets/jsrojas/ip-network-traffic-flows-labeled-with-87-apps},
  2017.
\newblock [Online; accessed 5-May-2023].

\bibitem{Irsoy2013}
O.~Irsoy, O.~T. Y{\i}ld{\i}z, and E.~Alpayd{\i}n, ``Design and analysis of
  classifier learning experiments in bioinformatics: survey and case studies,''
  {\em IEEE/ACM Transactions on Computational Biology and Bioinformatics},
  vol.~9, no.~6, pp.~1663--1675, 2012.

\bibitem{Shervin2019}
G.~Varoquaux and O.~Colliot, ``{Evaluating machine learning models and their
  diagnostic value},'' in {\em {Machine Learning for Brain Disorders}}
  (O.~Colliot, ed.), {Springer}, June 2023.

\bibitem{dataset}
B.~Rexha, A.~Musa, K.~Vishi, and E.~Martiri, ``Local media website network
  flows.''
  \url{https://github.com/ArbenaMusa/Local-Media-Website-Network-Flows}, 2023.
\newblock Accessed: September, 2023.

\end{thebibliography}

\end{document}